\documentclass[pre,aps,twocolumn,superscriptaddress,amsmath,showpacs,floatfix]{revtex4}
\usepackage{graphicx}
\begin{document}
\title{Membrane fluctuations around inclusions}
\author{Christian D. Santangelo}
\affiliation{Department of Physics, University of California, 
Santa Barbara, CA 93106} 
\author{Oded Farago}
\affiliation{Materials Research Laboratory, University of California, 
Santa Barbara, CA 93106}
\begin{abstract}
  \vspace{0.3cm} The free energy of inserting a protein into a
  membrane is determined by considering the variation in the spectrum
  of thermal fluctuations in response to the presence of a rigid
  inclusion.  Both numerically and through a simple analytical
  approximation, we find that the primary effect of fluctuations is to
  reduce the effective surface tension, hampering the insertion at low
  surface tension.  Our results, which should also be relevant for
  membrane pores, suggest (in contrast to classical nucleation theory)
  that a {\em finite}\/ surface tension is necessary to facilitate the
  opening of a pore.
\end{abstract}
\pacs{05.40-a; 87.16.Dg; 87.15.Kg}
\maketitle

Bilayer membranes are self-assembled thin fluid sheets of amphiphilic
molecules.  They are characterized by small bending and large
compression moduli, whose effective values are influenced by thermal
fluctuations ~\cite{renormalization}.  The softness of the bending
modes permit large shape deformations which are important for the
biological activities of some living cells (e.g., the red blood
cell)~\cite{lipowski}. Biological membranes are typically highly
heterogeneous: they usually consist of mixtures of different lipids
and, in addition, contain a variety of different proteins which carry
out diverse tasks such as anchoring the cytoskeleton, opening ion
channels, and cell signaling~\cite{cellbiology}.

Membrane inclusions can modify the thermal fluctuations of the
membrane by perturbing the local structure of the lipid matrix.  It is
well-known that the restrictions imposed on the thermal fluctuations
of the membrane are the origin of attractive van der Waals-like forces
between inclusions~\cite{entropicforces}.  While these long-range
interactions are typically very small, they are believed to play an
important role in determining the phase behavior (e.g. aggregation) of
such systems. Perturbing the spectrum of thermal fluctuations is also
expected to contribute to the free energy associated with the
insertion of proteins into lipid bilayers. This has an influence on
the solubility of proteins and other membrane inclusions. Remarkably,
this important entropic contribution to the insertion free energy of a
{\em single}\/ protein has been ignored in previous
calculations~\cite{previouscalcs}. In this letter we study the free
energy cost of inserting a rigid inclusion into a membrane, explicitly
taking into account effects due to membrane fluctuations. These
effects depend only on the inclusion's characteristic size. For
transmembrane proteins, the magnitude of the fluctuation free energy
can be as large as $25k_BT$. As this contribution is comparable, but
of opposite sign, to other free energy components, it strongly
influences the thermodynamic stability of proteins. Our results should
also be relevant for the fluctuation spectrum and nucleation energy of
a membrane pore. At low tension, the fluctuation free energy acts as a
barrier to the opening of a pore.

We consider a bilayer membrane consisting of $N$ lipids, that spans a
planar circular frame of a total area $A_p=\pi L_p^2$, in which a
rigid inclusion of radius $r_0\ll L_p$ has been inserted. The Helfrich
energy (to quadratic order) for a nearly-flat membrane in the Monge
gauge is given by~\cite{Helfrich}
\begin{equation}
{\cal H}_1 = \sigma A_p+\frac{1}{2} \int d^2\vec{r}\, 
\left[ \sigma \left(\nabla h \right)^2 + \kappa
  \left(\nabla^2 h \right)^2 \right],
\label{eq:helfrich} 
\end{equation}
where $\sigma$ is the surface tension, $\kappa$ the bending rigidity,
and $h$ the height of the membrane above the frame reference plane.
The boundaries of integration in Eq.(\ref{eq:helfrich}) include the
outer (frame, $r=L_p$) boundary and the inner (inclusion, $r=r_0$)
edge. The Laplacian in the Helfrich energy requires that we have two
boundary conditions (BCs) for each boundary. On the inner boundary we
fix the height of the membrane $h(r_0) = H(\phi)$ and the contact
slope $\partial h(r_0)/\partial r = H'(\phi)$, where $\phi$ is the
polar angle measured from the inclusion's axis of symmetry. On the
outer boundary will have the natural BCs: $h(L_p) = 0$ and $\nabla^2
h(L_p) = 0$. The particular choice of outer BCs does not modify the
free energy of the system in the thermodynamic limit.

To gain insight into the contribution of thermal fluctuations to the
insertion free energy we write the height function as $h = h_0 + f$
where $h_0$ is the extremum of Hamiltonian (\ref{eq:helfrich}), i.e.,
\begin{equation}
-\sigma \nabla^2 h_0+ \kappa \nabla^4 h_0 = 0,
\label{eq:h0}
\end{equation}
subject to the BCs that $h_0(r_0)= H(\phi)$, $\partial
h(r_0)/\partial r = H'(\phi)$, $h_0(L_p)= 0$, and $\nabla^2 h_0(L_p) = 0$.
This implies $f(r_0) = 0$ and $\partial f(r_0)/\partial r = 0$
on the inner boundary, and $f(L_p)= 0$ and $\nabla^2 f(L_p)=0$ on the
outer boundary. The Helfrich energy can be written as
\begin{eqnarray}\label{eq:helfrich2}
{\cal H}_1\left(h_0+f\right) = \sigma A_p+\int d^2\vec{r} \left\{ \frac{1}{2} 
\left[ \sigma \left(\nabla h_0 \right)^2 + 
\kappa \left(\nabla^2 h_0 \right)^2 \right]\right. \\
+\left[\sigma \nabla h_0 \cdot \nabla f 
+ \kappa \nabla^2 h_0 \nabla^2 f \right]
+  \left.\frac{1}{2}\left[ \sigma \left(\nabla f \right)^2 
+ \kappa \left(\nabla^2 f \right)^2 \right] \right\}\nonumber . 
\end{eqnarray}
For the cross term (third term in ${\cal H}_1$) we obtain, upon
integration by parts,
\begin{eqnarray}
&&\int d^2\vec{r} \left[\sigma \nabla h_0 \cdot \nabla f 
+ \kappa \nabla^2 h_0 \nabla^2 f \right]= \nonumber \\
&&\int d^2\vec{r}
\left[ - \sigma \nabla^2 h_0 + \kappa \nabla^4 h_0 \right]f
+ \int_{\partial M} \kappa \nabla^2 h_0 \left(\hat{n} \cdot \nabla\right) f + 
\nonumber \\
&&\int_{\partial M} \left( \hat{n} \cdot \nabla\right)\left[\sigma  h_0 
-\kappa  \nabla^2 h_0 \right] f, 
\label{crossterm}
\end{eqnarray}
where the last two integrals in the above equation are performed on
the boundaries of the system, and $\hat{n}$ is a unit vector normal to
the boundaries. The boundary terms in Eq.(\ref{crossterm}) vanish
since $f = 0$ and $\hat{n} \cdot \nabla f = -\partial f/\partial r=0$
on the inner boundary, and $f=0$ and $\nabla^2 h_0$ on the outer
boundaries. The bulk term also vanishes by virtue of Eq.(\ref{eq:h0}).

Without the cross term in Eq.(\ref{eq:helfrich2}), we are left with
three terms: the projected area term $\sigma A_p$, the equilibrium
term depending on $h_0$, and the fluctuation term depending on $f$.
Thus, the energies associated with $h_0$ and $f$ completely decouple
and their contributions to the free energy are additive. Note that in
our approach the equilibrium part of the free energy includes a
contribution from the height and tilt fluctuations of the {\em
  inclusion}\/. It is obtained by calculating the dependence of $h_0$
on the boundary values $H(\phi)$ and $H'(\phi)$, and performing an
appropriate thermal average over these quantities. Other energetic
components, such as hydrophobicity, translational entropy,
electrostatics, should be added to the equilibrium term, and can be
included in its definition \cite{abs}. The equilibrium term has been
analyzed in many previous studies~\cite{previouscalcs}. Its magnitude
is protein specific and is usually in the range of $-5$ to $-20k_BT$
\cite{lazaridis}. In contrast, the effect of {\em membrane}\/
fluctuations on the insertion free energy has not yet been considered
in the literature. We proceed to calculate the fluctuation part of the
insertion free energy. Note that it is independent of the height and
the contact angle of the inclusion (and their thermal fluctuations),
which affect only the equilibrium part \cite{remark}. 

Neglecting the equilibrium term, we are left with the projected area
and the fluctuation terms. By integrating the latter by parts twice,
the remaining Hamiltonian takes the form
\begin{equation} 
{\cal H}_2(f) = \sigma A_p+\frac{1}{2} \int
d^2\vec{r} f \left(-\sigma \nabla^2 + \kappa \nabla^4 \right) f.  
\label{helfrichfluct}
\end{equation} 
The boundary terms vanish in the above expression due to our choice of
BCs: $f(r_0)=0$, $\partial f(r_0)/\partial r=0$, $f(L_p)=0$, and
$\nabla^2 f(L_p)=0$. We proceed by expanding the function $f$ in a
series of eigenfunctions $f_{m,n}(r)$ of the operator ${\cal
  L}\equiv-\sigma \nabla^2 + \kappa \nabla^4$: $f(r,\phi) = \sum_{m,n}
h_{m,n}f_{m,n}(r) e^{i m \phi}$. The functions $f_{m,n}(r)$ can be
written as the linear combination of the Bessel functions, $J_m(r)$
and $Y_m(r)$, of the first and second kinds of order $m$, and the
modified Bessel functions of the first and second kinds of order $m$,
$K_m(r)$ and $I_m(r)$:
\begin{eqnarray} 
f_{m,n}(r)&=&A J_m(\lambda_1^{m,n} r) + B Y_m(\lambda_1^{m,n} r)\nonumber \\
&+& C K_m(\lambda_2^{m,n} r)  + D I_m(\lambda_2^{m,n} r), \nonumber
\end{eqnarray} 
where the $\lambda_i$ ($i=1,2$) are the positive solutions of
$(-1)^{i+1} \sigma (\lambda_i^{m,n})^2 + \kappa (\lambda_i^{m,n})^4 =
\mu_{m,n}$, and $\mu_{m,n}$ is the eigenvalue corresponding to the
function $f_{m,n}(r)$: ${\cal L}f_{m,n}(r)=\mu_{m,n}f_{m,n}(r)$.

Applying the BCs at $r_0$ and $L_p$, we derive the
eigenvalue equation
\begin{eqnarray}
& & \lambda_1 \left[ I_m(\lambda_2 r_0) K_m(\lambda_2 L_p) 
- I_m(\lambda_2 L_p) K_m(\lambda_2 r_0) \right]  \nonumber\\
& &\left[ Y'_m(\lambda_1 r_0) J_m(\lambda_1 L_p) 
- J'_m(\lambda_1 r_0) 
Y_m(\lambda_1 r_0) \right] \nonumber \\
& & = \lambda_2 \left[K'_m(\lambda_2 r_0) I_m(\lambda_2 L_p) 
- I'_m(\lambda_2 r_0) K_m(\lambda_2 L_p) \right] \nonumber\\
& &\left[J_m(\lambda_1 r_0) Y_m(\lambda_1 L_p) 
- J_m(\lambda_1 L_p) Y_m(\lambda_1 r_0) \right]
\label{eq:eigenvalues}
\end{eqnarray}
(for brevity, we have omitted the superscript $(m,n)$ from the
notation of the $\lambda_i$ in the above equation). In contrast, for
membranes without inclusions, we solve the simple equation
$J_m(\lambda_1 L_p) = 0$. It is interesting to note that, in the limit
that $\lambda_1^{m,n} r_0 \ll |m|$, Eq.(\ref{eq:eigenvalues}) reduces
to the eigenvalue equation in the absence of inclusions. This has the
physically appealing interpretation that modes with characteristic
lengths much larger than the inclusion radius are hardly perturbed by
its presence. In the opposite limit, $\lambda_1^{m,n} r_0 \gg |m|$
(which also implies $\lambda_1^{m,n} L_p \gg |m|$), we can neglect
terms proportional to $I_m(\lambda_i^{m,n} L_p)$ (which, otherwise,
become exponentially large) and replace the remaining Bessel functions
by their leading order asymptotic expressions. This gives, for
$\lambda_1^{m,n}\gg \sqrt{\sigma/\kappa}$, the simple equation $\tan
\left[\lambda_1^{m,n}\left(L_p-r_0\right)\right]= 1$, and the
solutions $\lambda_1^{m,n} \approx
\left[|m|/2+n+\left(-1\right)^m\pi/4\right]\pi/(L_p-r_0)$.  The
physical interpretation of this result is that the inclusion acts like
a hard wall for modes with characteristic lengths much smaller than
its radius. The effective linear size of the membrane for these modes
is reduced from $L_p$ to $L_p-r_0$ and the eigenvalues in this regime
increase by roughly a factor of $L_p/(L_p-r_0)$. Thus, the dominant
effect of the inclusion on the short wavelength modes is to lower the
density of contributing modes in ``$\lambda$-space'' [Note that
$\lambda_1^{m,n+1}-\lambda_1^{m,n}=\pi/(L_p-r_0)$].

When the function $f(r,\phi) = \sum_{m,n} h_{m,n}f_{m,n}(r) e^{i m
  \phi}$ is substituted in Hamiltonian (\ref{helfrichfluct}), we find,
due to the orthogonality the eigenfunctions
\begin{eqnarray}
&&\int_0^{2\pi}\!\!d\phi\int_{r_0}^{L_p} rdr\,
f_{m1,n1}(r)f_{m2,n2}(r)e^{i(m1+m2)\phi}\nonumber \\
&&=a_0\,\delta_{m1,-m2}\,\delta_{n1,n2},
\label{orthogonal}
\end{eqnarray}
that the modes decouple and that the Hamiltonian takes a quadratic
form in the amplitudes $|h_{m,n}|$. The normalization coefficient
$a_0$ in Eq.(\ref{orthogonal}) is the projected area per amphiphilic
molecule in the bilayer. Tracing over $|h_{m,n}|$ leads to the
following expression for the Gibbs free energy associated with
Hamiltonian ${\cal H}_2$ \cite{sens}
\begin{eqnarray}
\label{eq:gibbs}
&&G\left(\sigma,A_p\right)=\sigma A_p\\
&&+\frac{k_BT}{2} \sum_{m,n} \ln
\left\{\frac{\left[\sigma (\lambda_1^{m,n})^2 
+ \kappa (\lambda_1^{m,n})^4\right]A_p\lambda_{\rm dB}^2}
{2\pi k_BTN}\right\},
\nonumber
\end{eqnarray}
where $\lambda_{\rm dB}$ is the thermal de-Broglie wavelength of the
lipids.  The Helmholtz free energy is given by $F(A,A_p)=G-\sigma A$,
where the total membrane area $A$ is related to the surface tension by
\cite{remark2}
\begin{equation}
A\simeq A_p+\frac{k_BT}{2} \sum_{m,n} 
\frac{1}{\sigma+\kappa(\lambda_1^{m,n})^2}.
\label{areatension}
\end{equation}
Assuming that the membrane is incompressible and, therefore, that its
total area is fixed, we can use Eq.(\ref{areatension}) to derive the
following equation, relating the surface tension and the inclusion's
radius
\begin{equation}
-\pi r_0^2+\frac{k_BT}{2} \sum_{m,n}
\frac{1}{\sigma+\kappa(\lambda_1^{m,n})^2}-
\frac{1}{\sigma_0+\kappa(\lambda_{1,(0)}^{m,n})^2}=0.
\label{areatension2}
\end{equation} 
In the above equation $\lambda_{1,(0)}^{m,n}$ are the corresponding
solutions of the eigenvalue equation in the absence of the inclusion
($r_0=0$): $J_m(\lambda_{1,(0)}^{m,n}L_p)=0$, and
$\sigma_0\equiv\sigma(r_0=0)$. The solution to Eq.(\ref{areatension2})
has the form 
\begin{equation}
\sigma=\sigma_0(1+\delta),\ {\rm where}\ \ \delta\sim{\cal O}(r_0/L_p)^2.
\label{delta}
\end{equation}
The projected area and fluctuation parts of the insertion free
energy $\Delta F(r_0) \equiv F(r_0) - F(0)$ can now be calculated
using Eqs.(\ref{eq:gibbs}) and (\ref{areatension2}).
We find that $\Delta F(r_0)$ is given by
\begin{eqnarray}
\label{eq:freeenergydiff}
&&\Delta F(r_0) \approx - \pi \sigma_0 r_0^2 \\
&& +\frac{k_BT}{2} \sum_{m,n} \ln \left[\frac{\sigma_0 (\lambda_1^{m,n})^2 
+ \kappa (\lambda_1^{m,n})^4}{\sigma_0 (\lambda_{1,(0)}^{m,n})^2 
+ \kappa (\lambda_{1,(0)}^{m,n})^4}\frac{L_p^2-r_0^2}{L_p^2}
\right].\nonumber
\end{eqnarray}
Note that only $\sigma_0$ appears in the above expression, which is
due to Eq.(\ref{delta}) and the fact that we attempt to calculate
$\Delta F(r_0)$ only up to quadratic order in $r_0/L_p$. For the same
reason we can use $\sigma_0$ rather than $\sigma$ in the eigenvalue
equation (\ref{eq:eigenvalues}). The surface tension appears
implicitly in this equation, through the relation
$\lambda_2^2=\lambda_1^2+\sigma/\kappa$. In expression
(\ref{eq:freeenergydiff}) we assume that the number of molecules
forming the bilayer membrane does not change with the insertion of the
protein. Consequently, the total number of modes which is directly
proportional to the number of molecules in the bilayer is kept
constant. In contrast, the projected area per molecule [which appears
in Eq.(\ref{orthogonal})] does depend on the radius of the inclusion,
and this is the origin of the term $(L_p^2-r_0^2)/L_p^2$ appearing in
the argument of the logarithm in Eq.(\ref{eq:freeenergydiff}).  The
first term on the right hand side (r.h.s) of
Eq.(\ref{eq:freeenergydiff}) comes from the reduction of the projected
area. We will now show that, to a good approximation, the second term
on the r.h.s. of Eq.(\ref{eq:freeenergydiff}) is quadratic in $r_0$
and, thus, can be interpreted as a thermal correction to the surface
tension.

\begin{figure}[h]
\begin{center}
\scalebox{.4}{\centering \includegraphics{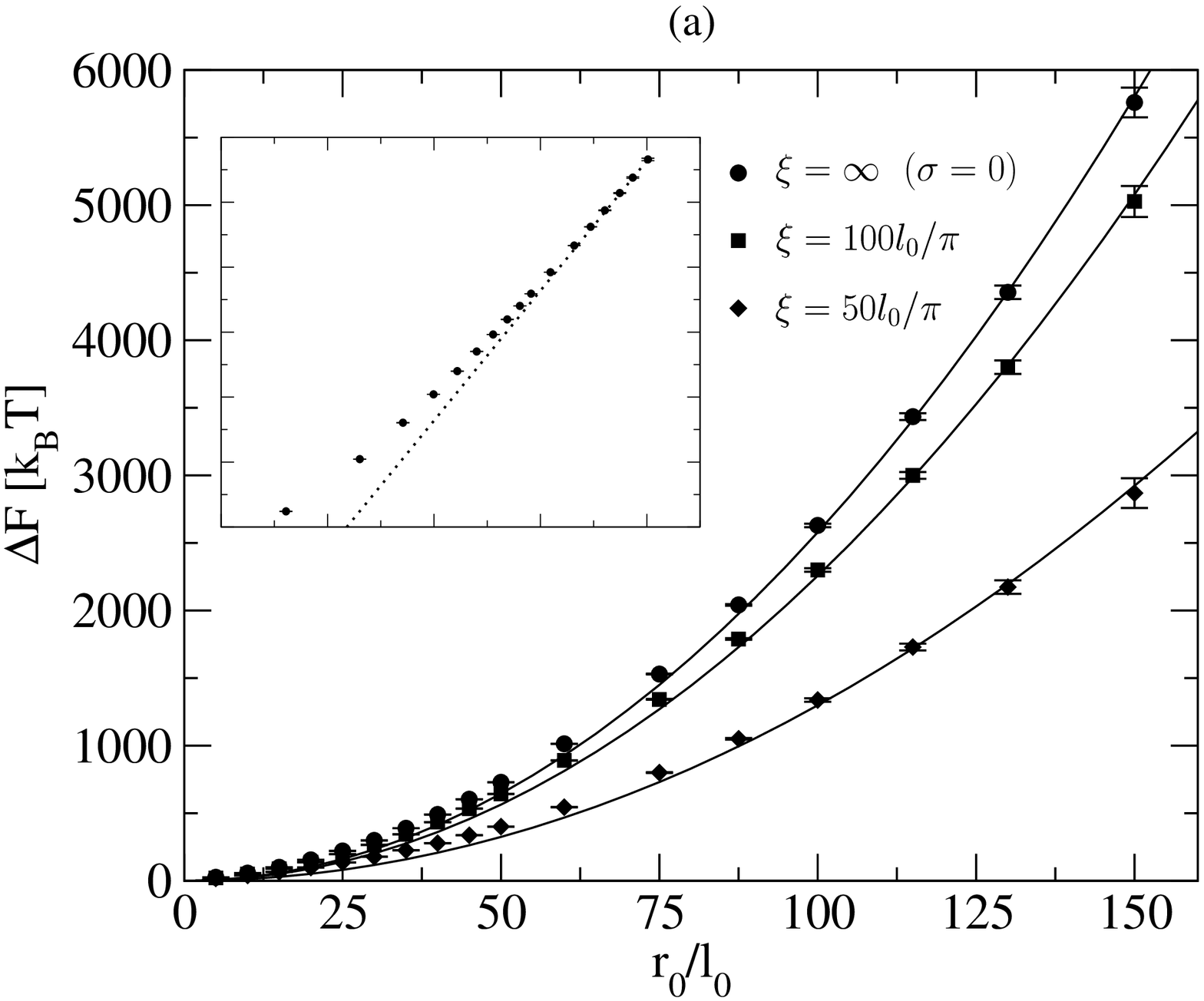}}
\vspace{0.3cm}
\scalebox{.4}{\centering \includegraphics{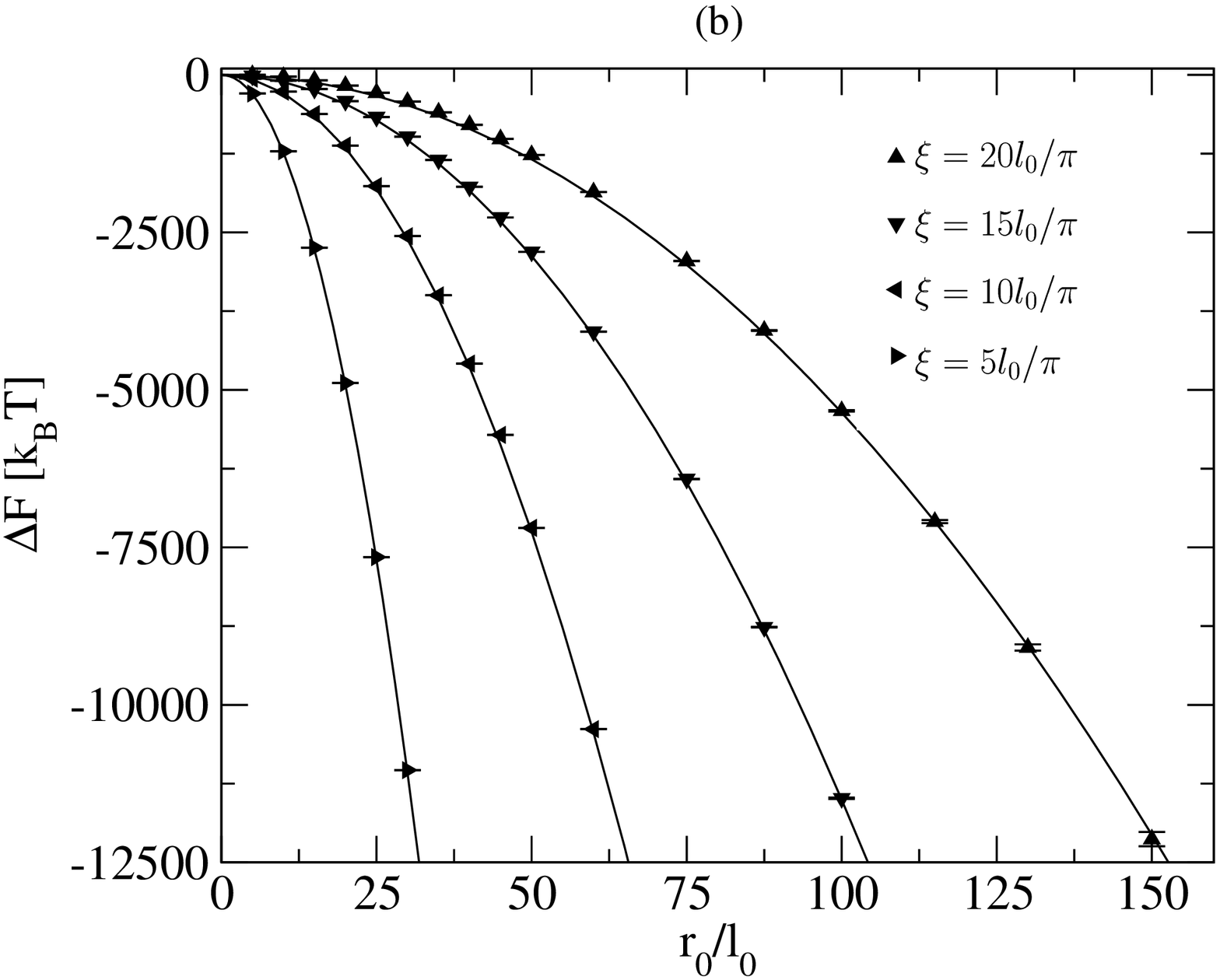}}
\end{center}
\vspace{-0.5cm}
\caption{The insertion free energy $\Delta F$ as a function of the 
inclusion's radius for $\kappa=10k_BT$ and various values of $\sigma_0$.
The inset to graph (a): a log-log plot of the numerical results for
$\sigma_0=0$. The slope of the straight dotted line is 2.}
\label{fig:fluct}
\end{figure}

In order to obtain an analytical result for the free energy
(\ref{eq:freeenergydiff}), we make the approximation [based on our
discussion of the asymptotic behavior of the eigenvalues
$\lambda_1^{m,n}$, see the text after Eq.(\ref{eq:eigenvalues})] that
eigenvalues such that $\lambda_1^{m,n} r_0 < \alpha |m|$ (long
wavelength) are not affected by the inclusion, whereas modes with
$\lambda_1^{m,n} r_0 > \alpha |m|$ (short wavelength) grow by a factor
$L_p/(L_p-r_0)$. The numerical constant $\alpha$ is of the order of
unity and its value, which may depend on the surface tension
$\sigma_0$, will be determined later by an exact numerical evaluation
of $\Delta F$.  We have verified numerically that this asymptotic
form is indeed correct. We set $n=0,1,\ldots,\sqrt{N_0}$, and,
$m=-\sqrt{N_0},\ldots,\sqrt{N_0}$ so that the total number of modes
(degrees of freedom), $2N_0$, is proportional to the number of
molecules forming the membrane, $N$. Along with these approximations,
we evaluate the sum in equation (\ref{eq:freeenergydiff}) as an
integral, giving us the simple result (correct up to quadratic order
in $r_0$) that $\Delta F = - \pi(\sigma_0-\sigma^*)r_0^2$, where
\begin{equation}
\sigma^* = \frac{k_B T}{\pi\alpha\ell_0^2}  
\left\{ 2 - \alpha-\left( \frac{\ell_0}{\pi\xi}\right)^2 
\ln \left[ \left(\frac{\pi\xi}{\ell_0}\right)^2 + 1 \right] \right\},
\label{rentension}
\end{equation}
$\xi=\sqrt{\kappa/\sigma_0}$, and $\ell_0=L_p/\sqrt{N_0}$ is a
microscopic length cutoff on the order of the characteristic size of a
membrane molecule. We thus obtain the result that the fluctuations
renormalize the surface tension. It is interesting to note that this
renormalization tends to occur with the opposite sign as the bare
surface tension (for $\ell_0 \lesssim \xi$), thus making it
\textit{harder}\/ to insert an inclusion. Only for very stressed
membrane ($\xi \lesssim \ell_0$) does $\sigma^*$ become negative.
This is due to the reduction of the projected area that allows more
thermal fluctuations. A more careful analysis of the long wavelength
modes shows that these contribute only finite-size effects to the free
energy which vanish in the limit of $L_p \gg r_0$.

We have numerically solved the eigenvalue equation
(\ref{eq:eigenvalues}) and used the solutions to evaluate the sum in
Eq.(\ref{eq:freeenergydiff}). Numerical values of $\Delta F(r_0)$ (for
$\kappa=10k_BT$ and various values of $\sigma_0$) are shown in
Fig.\ref{fig:fluct} (a)-(b). They have been extracted by extrapolating
the numerical results obtained for several values of $750\leq N_0\leq
2000$ to the thermodynamic limit $N_0\rightarrow\infty$. In the inset
to Fig.\ref{fig:fluct} (a), the results for $\sigma_0=0$ are replotted
on a logarithmic scale, showing that our prediction of a quadratic
relation between $\Delta F$ and $r_0$ is attained only for large
inclusions with $r_0\gtrsim 100\ell_0$ (the slope of the straight
dotted line is 2). This is a typical size for colloidal particles
\cite{koltover}. The value of the constant $\alpha$ appearing in
Eq.(\ref{rentension}) shows a slight dependence on the surface tension
varying from 1.59 for $\sigma_0=0$ to 1.72 for
$\xi=\sqrt{\kappa/\sigma_0}=5l_0/\pi$. The solid curves in
Fig.\ref{fig:fluct} (a)-(b) depict our analytical expression for
$\Delta F$, with $\alpha$ determined by fitting the results for large
$r_0$ to Eq.(\ref{rentension}). From Fig.\ref{fig:fluct} (a) we
conclude that, because of thermal fluctuations, there is a free energy
penalty to embedding an inclusion in a weakly stretched membrane
(small $\sigma_0$). For transmembrane proteins with typical radii of
$r_0\lesssim 5\ell_0$, the energy cost is $\Delta F\lesssim 25k_BT$,
which is comparable to the equilibrium contribution but of opposite
sign. This demonstrates the importance of the membrane fluctuations in
determining the distribution of transmembrane and free proteins. For
larger inclusions, the fluctuation free energy will dominate the
equilibrium part. On the other hand, Fig.\ref{fig:fluct} (b) shows
that inclusions greatly reduce the free energy of strongly stretched
membranes (large $\sigma_0$). The primary reason that the free energy
is lowered in this regime is the reduction of the projected area.
These results should also be relevant for the question of nucleation
of a membrane pore which, albeit more complicated, can be studied by
similar approach~\cite{us}.  They suggest that there exists a
(finite!)  critical value of the surface tension below which pores
cannot open and above which they grow without bounds.  Classical
nucleation theory, which ignores fluctuations effects, predicts that
the critical surface tension is zero~\cite{nucleationtheory}.

In summary, we have computed the free energy of inserting an inclusion
into a membrane. We explicitly calculated the contribution of membrane
fluctuations. The primary effect of these fluctuations is to reduce
the effective value of the surface tension. At low surface tension it
provides a positive component to the free energy of an embedded
inclusion, thereby impeding the insertion of transmembrane proteins.
The sensitivity of the free energy to variations of the surface
tension suggests that, by controlling the membrane surface tension
appropriately, one may control the thermodynamic stability of embedded
proteins and, thus, the equilibrium distribution between proteins
inserted in the membrane and in solution.

{\em Acknowledgments:}\/ We thank M. Kardar, A.W.C. Lau, and P.
Pincus for useful discussions.  This work was supported by the NSF
under Award No. DMR-0203755.  The MRL at UCSB is supported by NSF No.
DMR-0080034.


\begin{thebibliography}{99}
\bibitem{remark} The inner boundary reflects the projection of the
  cross-sectional area of the inclusion onto the frame reference
  plane. In the above derivation, we consider a circular boundary with
  a fixed radius $r=r_0$. However, the locus of the inner boundary
  depend on the tilt angle of the inclusion and varies accordingly.  A
  straightforward calculation {\protect \cite{us}} shows that if the
  tilt angle is small (i.e., when the inner boundary only slightly
  deviates from circularity) then the boundary of integration in
  Eq.(\ref{eq:helfrich}) can be still taken as circular at the expense
  of introducing an additional boundary term in the Hamiltonian. The
  new boundary term has no influence on the membrane fluctuations
  which are governed by the surface Hamiltonian only. Therefore, and
  for the sake of the simplicity of our derivation, we have dismissed
  this extra boundary term and its derivation from the discussion.
\bibitem{remark2} Corrections to this relation can be neglected when
  the tilt angle and height of the membrane on the inner boundary are
  small: $H'\ll 1$, $HH'\ll r_0$.
\end{thebibliography}
\end{document}